\documentclass{article}

\usepackage{graphicx}
\usepackage{natbib}
\usepackage{amsmath} 
\usepackage{amssymb}
\usepackage{graphicx}
\usepackage{hyperref}
\usepackage{color}
\usepackage{multicol} 

\usepackage{multirow} 
\usepackage{makecell} 

\newtheorem{thm}{Theorem}[section]

\newcommand{\re}{\mathbb R}
\newcommand{\norm}[1]{\left\lVert#1\right\rVert}
\DeclareMathOperator{\ima}{Im}
\DeclareMathOperator{\logit}{logit}

\begin{document}

\title{Generalized Linear Models for Geometrical Current predictors. An application to predict garment fit}
\author{ Sonia Barahona$^{(1)}$, Pablo Centella$^{(1)}$, Ximo Gual-Arnau$^{(2)}$,  Maria Victoria Ib\'a\~nez $^{(3)}$ and Amelia Sim\'o$^{(3)}$ \\
        \small{(1)     Department of Mathematics. Universitat Jaume I.
              Avda. del Riu Sec s/n. 12071-Castell\'on, Spain.}
                           \\
          \small{(2)   Department of Mathematics-INIT. Universitat Jaume I.
              Avda. del Riu Sec s/n. 12071-Castell\'on, Spain.}
                                      \\
         \small{(3)    Department of Mathematics-IMAC. Universitat Jaume I.
              Avda. del Riu Sec s/n. 12071-Castell\'on, Spain.}
    }

\maketitle

\begin{abstract}
The aim of this paper is to model an ordinal response variable in terms of vector-valued functional data included on a vector-valued RKHS.
In particular, we focus on the vector-valued RKHS obtained when a geometrical object (body) is characterized by a \emph{current} and on the ordinal regression model.
A common way to solve this problem in functional data analysis is to express the data in the orthonormal basis given by decomposition of the covariance operator. But our data present very important differences with respect to the usual functional data setting. On the one hand, they are vector-valued functions, and on the other, they are functions in an RKHS with a previously defined norm. We propose to use three different bases: the orthonormal basis given by the kernel that defines the RKHS, a basis obtained from decomposition of the integral operator defined using the covariance function, and a third basis that combines the previous two. The three approaches are compared and applied to an interesting problem: building a model to predict the fit of children's garment sizes, based on a 3D database of the Spanish child population.

\textbf{keyword} Statistical Shape and Size Analysis; Vector-valued Reproducing Kernel Hilbert Space; Functional Data Analysis; Ordinal Regression.

\end{abstract} 

\section{Introduction} 

In many scientific fields, such as Biology, Medicine and Anthropometry, we can find a great number of applications where it is necessary to predict a categorical variable as a function of a geometrical object predictor. These geometrical objects can be mathematically characterized in different ways, the most popular being as a set of landmarks (\cite{Bookstein78,Kendall84,dryden2016statistical}), compact sets (\cite{Serra82,BaddeleyMolchanov98,simo2004resuming,molchanov2006theory}) or functions (\cite{loncaric1998survey,Kindratenko,gual2013shape}).
In this paper, the contour of each geometrical object (surface in $\re^3$) is represented by a mathematical structure named current. This framework was introduced by \cite{VaillantGlaunes05} and \cite{Glaunesetal06} and it provides a unifying framework in which to process any sets of points, curves and surfaces or a mixture of these. No hypothesis on the topology of the shapes is assumed. Moreover, it is weakly sensitive to the sampling of shapes and it does not depend on the choice of parameterizations.

Currents are mathematically complex objects but, fortunately, it is possible to associate a subspace of currents with a vector-valued Reproducing Kernel Hilbert Space (RKHS) by duality and, as a result, we can represent each geometrical object with a function in an RKHS (\cite{Durrleman10, Barahonaetal16}). A vector-valued RKHS is a Hilbert vector space of functions with useful properties.

\vspace{0.2cm}

This work is motivated by an experimental study carried out by the Biomechanics Institute of Valencia, whose ultimate objective was to implement a web application for online shopping for children's wear.
In particular, that application should make it possible to select the right size of children's clothing without requiring the child to try on the clothes.
Selecting the proper size of any garment for a child without trying it on constitutes a problem when buying these items both in a physical store and, especially, online.
Seventy-eight randomly selected children between the ages of $3$ and $12$ years participated in this study. Firstly, the children were scanned using a 3D body scanner. Next, garments were tried on in different sizes and an expert classified the fit of each garment as ``too small", ``good fit" or ``too big".

\vspace{0.2cm}

Different approaches to this problem can be found in the literature. Most of them are based on taking the user's anthropometric measurements and their relationship with the dimensions of the garment. In this paper, we will describe a method based on 3D scanning of the child's body. Using the currents approach, a child will be represented by a function in a vector-valued RKHS and Functional Data Analysis (FDA) will be used on this space.
Unlike the methods based on landmarks, curves or parameterized surfaces, the great advantage of working with functions is that the shift from two to three dimensions does not increase the complexity of the expressions or the calculations.

Nowadays, the theory of statistics with functional data is an important field of research in statistics. It is used when data are in an infinite-dimensio\-nal function space. Although this theory is often a generalization of classic parametric or multivariate statistics, the infinite-dimensional nature of the sample space poses particular problems. Key references in the FDA literature are the books by \cite{SilvermanRamsay05} and \cite{FerratyVieu06}. A more theoretical treatment is taken in \cite{Hsing15}.

\vspace{0.2cm}

With respect to the particular problem of regression with a scalar response and a functional predictor, the first papers focused on the continuous version of the multiple linear model, the functional linear model. In these cases, a direct estimation of the parameter function of the proposed functional regression models through the use of least squares methods is not possible. The most commonly used approximated solution for this estimation problem is to consider that functional observations belong to a space generated by a basis of functions and to perform a multiple treatment based on this approach. Different bases have been used in the literature, such as spline functions, trigonometric functions or wavelet functions (\cite[see][and the references therein]{SilvermanRamsay05}).

A slightly different approach, and one of the most popular, is to use principal component functional regression. This approach uses the orthonormal basis of eigenfunctions of the covariance function (\cite{cardot99}). Unlike the previous ones, this is a data-driven basis. As in the multivariate case, this technique makes use of the data covariance function to determine the subspace where the data are projected. This subspace is spanned by the data covariance eigenfunctions and it is always an RKHS.
This approach solves a typical problem in functional regression: the great dependence between coefficients, that causes that the estimation of the model is not very accurate.

Nevertheless, as \cite{morris2015} advises, care must be taken when using principal component functional regression for very complex, high-dimensional functional data for which the decay rate in the eigenvalues is slow, especially when the number of functions is small. In certain high-dimensional, low-sample-size settings, PCs have been shown to be inconsistent.
These problems can be at least partially mitigated using functional principal component analysis with regularization (\cite{Yuan10}).

\vspace{0.2cm}

In the same spirit as in the multivariate setting, functional generalized linear models (\cite{escabias2004principal}), which are the functional version of generalized linear models (\cite{NelderWedderburn72,MccullaghNelder89}), were introduced more recently in the literature. Such models are based on similar ideas to the linear case. In \cite{James02} the predictors are modeled as cubic splines, and in \cite{cardotetal05} the functional coefficient of the functional generalized linear model is estimated via penalized likelihood with spline approximation. \cite{dou12} use the functional principal component analysis (FPCA) approach.

However, when we use the FPCA approach in logistic or multinomial functional generalized linear models, we face an additional problem to the one mentioned previously. Logistic or, in general, multinomial generalized linear models, are used to solve classification problems and, as explained in \cite[chap. 9]{jolliffe02}, when PCA is used in classification problems in order to reduce the dimensionality of the analysis, we have to be aware that there is no guarantee that the separation between groups will be in the direction of the high-variance PCs; the separation between groups may be in the directions of the last few PCs.

On the other hand, as \cite[chap. 3]{Gonzalez10} notes, an improvement in the classification results can be achieved by using other kernels that capture nonlinear dependencies between the data, because the covariance function only deals with linear ones.

 RKHSs (usually, scalar-valued) have a long history in the statistical and machine learning literature.  RKHSs have been largely used to facilitate statistical modeling and estimation. For example, in the 1940s, probabilists had already begun to employ Hilbert space methods to clarify the structure of time series (\cite{parzen1961approach}).  \cite{preda2007regression, Yuan10} and \cite{cai2012minimax} use the RKHS framework in prediction problems under which the unknown slope function is assumed to reside in a reproducing kernel Hilbert space $H(K)$ but originally data are in $L^2$ with the usual inner product. In the literature of Support Vector Machines, RKHSs are used to map original data in a higher dimensional space (\cite{cristianini2000introduction,steinwart2008support}). RKHSs provide a convenient framework for
efficient computation.

An important and crucial difference between previous applications of RKHSs and ours is that, in our case, original data are functions in a given vector-valued Reproducing Kernel Hilbert Space with a previously defined inner product, that is generally different from the $L^2$ inner product. For example, polynomial functions can never be included in an RKHS with a Gaussian kernel (\cite{steinwart2008support}).

Because our functional data are vector fields in an RKHS, we propose three different bases to express our vector fields with respect to them. First, the orthonormal basis given by the integral operator defined by its reproducing kernel(~\cite{Quangetal10}). Secondly, we consider the integral operator from the covariance function regarding our data as a realization of a stochastic process. The relationship between $L^2$, our original RKHS and the RKHS defined from the covariance function provides another non-orthonormal base.
Finally, we prove a result on simultaneous diagonalization that provides an alternative basis system that combines
the properties of both.
The three approaches are compared in a very novel application to online clothing sales.

Our implementations have been written in~\cite{Matlab15} and R (~\cite{R2017}).

The article is organized as follows:
Section~\ref{MotivatingExample} provides a detailed explanation of the practical case that motivated this work.
Section~\ref{sec:RKHS} introduces the concepts of currents and Reproducing Kernel Hilbert Spaces.
Section~\ref{sec:FGLM} reviews the basis of functional generalized linear models and Section~\ref{sec:base-ortonormal} gives the different bases of functions in the RKHS.
The application for predicting children's garment fit is detailed in Section~\ref{sec:our_appl}.
Finally, conclusions are discussed in Section~\ref{sec:conclusions}.

\section{Motivating Example}\label{MotivatingExample}

In the current process of buying children's clothing online, consumers base the size selection on their previous experience or on the size chart that is normally included in the online store. The consumer's previous experience is not usually very reliable, because each brand uses its own sizing system
that usually evolves over time according to the needs of each company. Size tables indicate the ranges of the main anthropometric measures covered by each size. This method is also unreliable, because taking measurements at home is subject to significant errors, and ambiguous, because users can fit into different sizes according to the measurements used. The final result is a high percentage of returns on children's clothing sold online, meaning that many consumers are reluctant to buy through this channel, thus increasing the cost of sales.

Selection based on the child's anthropometry seems the most appropriate approach to predict garment size and fit in the child population. However, the 3D anthropometry acquisition systems currently available have several drawbacks. 3D body scanners are too expensive for home use. Parametric avatars configured from manual measurements by the user present three important sources of inaccuracy: they are not based on statistics of real populations but on models and proportions that prevail in aesthetics, the number of measurements entered does not depend on the type of garment selected or the critical measurements for the associated adjustment, and the measurements are taken by an untrained user, with the error that this may entail. Low-cost systems that use domestic technology to capture body measurements have not yet achieved sufficient precision for size allocation or prediction of fit.

Ergonomic childrenswear  design and size definition processes have several differences with regard to those of adult apparel. Firstly, childrenswear size designation is usually labeled in ages, which is not a body measurement, so it is usually related to a specific body height per age, which may not necessarily be close to a child of that age, due to the high variability of height by age in children.
According to the European standard UNE-EN 13402-3, the 3 to 12 years age range has 10 different sizes associated with it (950-1010 mm, 1010-1070 mm, 1070-1130 mm, 1130-1190 mm, 1190-1250 mm, 1250-1310 mm, 1310-1370 mm, 1370-1430 mm, 1430-1490 mm and 1490-1550 mm).

As online clothes shopping is a problem for both the customer and the apparel industry, in recent years both national administrations and industrial groups from the clothing sector have been developing national anthropometric surveys in different countries. Emerging technology for body scanning has also promoted these new sizing surveys.

In order to help solve all these problems, the Biomechanics Institute of Valencia (IBV) started an ambitious research project in 2004, of which this work is a part. This project has two objectives: first, to develop a system for capturing the child body's 3D morphometry that is precise, easy to use and can be done at home. Second, to build a model to predict how a given garment size of garment fits a child based on the aforementioned 3D reconstruction. Our work in this article focuses on the latter.

With respect to the first objective of the project addressed by the IBV group, an application has already been developed.  The system reconstructs the body of the child in 3D from two or three photographs taken with domestic technology (smartphone, tablet or digital camera) using models representative of the European infant morphometry as a base for reconstruction~\citep{ballesteretal16}.

To achieve both objectives, a 3D anthropometric study of the child population in Spain was conducted in 2004.
In this study, a randomly selected sample of Spanish children between the ages of $3$ and $12$ years was scanned using a Vitus Smart 3D body scanner from Human Solutions, a non-intrusive laser system which performs a sweep of the body. Several cameras capture images and associated software provided by the scanner manufacturer provides information about the 3D spatial location of up to $200000$ points on the body surface.
3D scan data was processed for the creation of posture harmonized homologous models to obtain a database of individual 3D homologous avatars with anatomical one-to-one vertex correspondence among them. Next all the scans were rigidly aligned(~\cite{ballesteretal14}).

Seventy-eight of these children of different ages performed an additional fit test, where they tested up to three different consecutive sizes of the same shirt model: the supposedly correct size, the size above and the size below. Then, an expert in clothing and design evaluated each fit qualitatively (as small, correct fit or large).  There were 7 possible shirt sizes available, nominally corresponding to ages 3, 4, 5, 6, 8, 10 and 12.  In 24 cases, only two sizes were evaluated.  In 18 of these cases, the children tested either had a correct shirt size corresponding to ages 3 or 12. The 6 remaining  cases with only two sizes evaluated were due to lack of cooperation by the children. Additionally, 9 children tested just one shirt size because the age 12 size was too small for them or the age 3 size was too large for them.

So, our data set contains the 3D body scans of a total of 78 children (37 boys and 41 girls, between 3 and 12 years old). It also includes  the expert's opinion on the goodness of the fit  of different (consecutive) sizes of the same shirt model on the children, codifying the goodness of fit as -1 (if the shirt is too small), 0 (for a good fit) or 1 (if the shirt is too big). The total number of expert observations is 192 (3 evaluations of 45 children, 2 evaluations of 24 children and 1 evaluation of 9 children).

\begin{figure}
\begin{center}
\begin{tabular}{c}
\includegraphics[width=4.5cm]{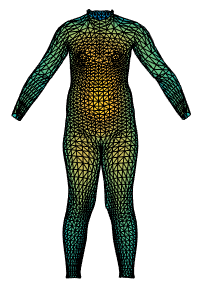}
\end{tabular}
\caption{Triangulated surface from a child's scan from the database. \label{nene}}
\end{center}
\end{figure}

As the children's head, hands, legs and feet do not come into play in shirt size selection, these parts were discarded from the scans, and a total of $1423$ points representing the remaining surface per child were considered.
This amount of detail is enough to characterize a child for our purposes, while keeping the time and memory requirements to perform the calculations reasonable. These points were grouped into $2766$ triangles forming a mesh.
The body contour from each child in our data set was therefore represented by an oriented triangulated smooth surface, $S_k$  (see Fig.~\ref{nene}).

\section{3D geometrical objects as elements in a Reproducing Kernel Hilbert Space} \label{sec:RKHS}

Currents are mathematical elements that can be used to model general geometrical objects(~\cite{Durrlemanetal09,Barahonaetal16}). In this paper, they will be used to model the bodies of the children in our data set.

Let $D$ be a compact set in $\re^3$ and $K:D\times D\longrightarrow \re^{3\times 3}$, the matrix-valued kernel associated with a vector-valued RKHS $H_K (D,\re^3)$.

The current representation of a surface $S \subset D$ is defined by the integral of $K$ along the surface:

\begin{equation}
C_S(y)=\int_S K(x,y) ( \tau (x)) \,dx. \nonumber
\end{equation}
Where $\tau(x)$ is the normal vector to the surface $S$ at point $x$(~\cite{Durrlemanetal09, Barahonaetal16}).

In the discrete setting, the vectors $\tau(x)$ are constant over each mesh cell. Then, if $x_j$ is located at the center of mass of mesh cell $j$, and $\tau_j$ is $\tau(x_j)$ scaled by the size of the mesh cell,
\begin{equation}\label{eq:discrete}
S \longrightarrow C_S(\omega)\cong \int_S K(x, \cdot) (\tau(x))\,dx\approx \sum_j K(x_j, \cdot) (\tau_j),
\end{equation}

So, the vector field  $\varphi_k$ associated  with each surface $S_k$ will be defined on a different set of points $\{x_k\}$ given by the centers of mass of
the respective mesh cells. Using the ``Representer Theorem" (\cite{CuckerSmale01}),  given $\{a_i\}_{i=1}^N$ a sample grid in D,
we can find a smooth function $\overline{\varphi_k}$, defined as:
\begin{equation}\label{eq:funcio7}
\overline{\varphi_k}= \sum_{i=1}^N K(a_i, \cdot)(\beta_i^k),
\end{equation}
where $\overline{\varphi_k}(a_i)$ is closest to $\varphi_k(a_i)$ (~\cite{Barahonaetal17}).

The inner product of two geometrical objects represented as currents is given by the inner product of the corresponding elements in $H_K (D,\re^3)$.;  that is, if
$\overline{\varphi_1}=\sum_j K(a_j^1, \cdot) (\beta_j^1)$ and $\overline{\varphi}_2=\sum_j K(a_j^2, \cdot) (\beta_j^2)$ are two elements in the RKHS, associated with two surfaces $S_1$ and $S_2$,
then $$\langle \overline{\varphi_1}, \overline{\varphi_2} \rangle_{H_K}=\sum_j\sum_l \beta_j^1 \cdot K(a_j^2, a_l^1) (\beta_j^2),$$
where $\cdot$ is the inner product in $\re^3$.

From now on, the vector fields in $H_K(D, \re^3)$ will be named functions and $H_K(D, \re^3)$ will be denoted $H_K$ for the sake of simplicity.
The space of quadratic integrable vector fields from $D$ to $ \re^{3}$ with the Lebesgue measure will simply be denoted by $L^2$.

Moreover, a Gaussian kernel, $K(x,y)=k(x,y) I_{3\times 3}:= \exp(\frac{-\norm{x-y}^2_{\re^3}}{\lambda^2})I_{3\times 3}$ will be used in the definition of the operator-valued reproducing kernels $K$, where $I_{3\times 3}$ denotes the identity matrix (\cite{Barahonaetal17}).

\section{The functional generalized linear model} \label{sec:FGLM}

As mentioned in the introduction, the aim of this paper is to predict the fit of a particular garment on a child as ``too small" ($Y=-1$), ``good fit" ($Y=0$) or ``too big" ($Y=1$), given different predictors, including the child's surface, which, as shown in the previous section, is modeled as a current.

Generalized linear models (GLMs)(~\cite{MccullaghNelder89}) provide a natural generalization of classical linear models. Given a dependent variable $Y$, with $E(Y)=\mu$, they assume that it is distributed following a probability function in the exponential family (not necessarily a Gaussian distribution),
and the relationship between predictors $\varphi_1,\dots,\varphi_m$ and response $Y$ is modeled by means of a link function $g$ as:
\begin{equation}\label{eq:funcionlink}
g(\mu)=\alpha +\sum_{i=1}^m \sigma_i \varphi_i
\end{equation}
where $g$ can be any monotonic differentiable function, and $\alpha$ and $\sigma= (\sigma_1,\dots,\sigma_m)$ are the parameters to estimate.

When the response variable is a score, representing an ordered category, i.e. when $Y$ may take one of several discrete ordered values indexed as $1,\dots,J$ with probabilities $\pi_1,\dots,\pi_J$: $\sum_{j=1}^J \pi_j=1$, these probabilities can be modeled using cumulative logits (~\cite{agresti10}) as:

\begin{equation}\label{eq:modelo-logit}
\logit[P(Y \leq j)]=\alpha_j +\sum_{i=1}^m \sigma_i \varphi_i, \hbox{\hspace*{0.8cm}}  \forall j\in \{1, \dots, J-1\}.
\end{equation}

GLMs provide a very flexible class of procedures. However, they assume that the predictors have a finite dimension. That is why~\cite{James02} extended GLMs to functional generalized linear models (FGLMs), which directly model the relationship between a single response from any member of the exponential family of distributions and a functional predictor. Then, when the predictor $\varphi(\cdot)$ is functional, as in our case, the link given by Eqs.~\eqref{eq:funcionlink} and \eqref{eq:modelo-logit} cannot be applied directly, but a natural generalization is to replace the summation over the finite-dimensional space with an integral over the infinite-dimensional one. Then,
if we were working with functional predictors in $L^2$, the cumulative logit model (Eq.~\ref{eq:modelo-logit}) would become:
\begin{equation*}
\logit[P(Y \leq j)]=\alpha_j +\int \sigma(x) \varphi(x) \,dx, \forall j\in \{1, \dots, J-1\},
\end{equation*}
where $\sigma(\cdot)$ is the functional analogue of $\sigma$ in Eq.~\eqref{eq:modelo-logit}.
But, in our case, we have functional predictors in $H_K$, so:
\begin{equation}\label{eq:modelo-logit2}
\logit[P(Y \leq j)]=\alpha_j + \langle \sigma, \varphi \rangle_{H_K}, \forall j\in \{1, \dots, J-1\}.
\end{equation}

The most widely used approach to estimate these models considers that the functions $\sigma(\cdot)$ and $\varphi(\cdot)$ belong to spaces generated by bases of functions
(\cite[see][]{SilvermanRamsay05}). In our case, $\sigma(\cdot)$ and $\varphi(\cdot)$ belong to a common space (a vector-valued RKHS),
so if $\{\phi_l(\cdot)\}_{l=1}^{\infty}$ is a basis of this space, there will exist coefficients $\{c_l\}_{l=1}^{\infty}$, $\{b_l\}_{l=1}^{\infty}$ such that:
\begin{equation*}
\varphi(x)=\sum_{l=1}^\infty c_l \phi_l(x); \hbox{   \hspace*{2cm}  } \sigma(x)=\sum_{l=1}^\infty b_l \phi_l(x).
\end{equation*}

In practice, these developments are truncated, and $\varphi(x)$ and $\sigma(x)$ are usually approached by a summation of a finite number of terms, as:
\begin{equation}\label{eq:bases-de-funciones}
\varphi(x)\cong \sum_{l=1}^r c_l \phi_l(x); \hbox{   \hspace*{2cm}  }  \sigma(x) \cong \sum_{l=1}^r b_l \phi_l(x).
\end{equation}

Ideally, these basis functions should have similar features to the functions being estimated.
Different bases have been used in the literature, such as trigonometric functions, spline functions(~\cite{Aguilera96}), wavelet functions(~\cite{Ocana98}) or the orthonormal basis of eigenfunctions of the covariance function(~\cite{cardot99}).

Then from Eqs.~\eqref{eq:modelo-logit2} and \eqref{eq:bases-de-funciones}, the link function of the cumulative logit models can be written as:
\begin{equation}\label{eq:modelo-logit3}
\logit[P(Y \leq j)]=\alpha_j +\sum_{p=1}^r \sum_{l=1}^r  b_p c_l \langle \phi_p, \phi_l\rangle_{H_K},\hbox{   \hspace*{0.5cm}  }  \forall j\in \{1, \dots, J-1\}.
\end{equation}

And then:
\begin{equation}\label{eq:logit-final}
P(Y \leq j)=\frac{\exp\biggr(\alpha_j +\sum_{p=1}^r \sum_{l=1}^r  b_p c_l \langle \phi_p, \phi_l\rangle_{H_K} \biggr)}{1+\exp\biggr(\alpha_j +\sum_{p=1}^r \sum_{l=1}^r  b_p c_l \langle \phi_p, \phi_l\rangle_{H_K}\biggr)}, \ \forall j\in \{1, \dots, J-1\},
\end{equation}
where $\{\alpha_j\}_{j=1}^{J-1}$ and $\{b_l\}_{l=1}^r$ are the parameters to estimate.

If the basis is orthonormal for $H_K$, then Eq.~\eqref{eq:logit-final} becomes:
\begin{equation*}
P(Y \leq j)=\frac{\exp\biggr(\alpha_j + \sum_{l=1}^r  b_l c_l  \biggr)}{1+\exp\biggr(\alpha_j + \sum_{l=1}^r  b_l c_l \biggr)}, \qquad \forall j\in \{1, \dots, J-1\}.
\end{equation*}

\section[Bases of functions in \$H\_K(D, R\textasciicircum n)\$]{Bases of functions in $H_K$} \label{sec:base-ortonormal}
This section focuses on the methodological novelties of our work.
It is interesting to remember at this point that we are working in a vector-valued RKHS, i.e. our data are vectorial fields.
Theoretical properties about bases in scalar RKHSs are well known and scalar RKHSs and their bases have been largely used in the statistical literature. However, nowadays theoretical properties of vector-valued RKHSs in general and their bases in particular are a research field in the functional analysis literature and, as far as we know, they have never been used in classical statistical applications.
Vector-valued RKHSs were used in Image colorization problems, a particular case of a mathematical extension problem(~\cite{Quangetal10}).

It should be noted that in conventional functional data analysis applications, the original data are functions in the $L^2$ Hilbert space and RKHSs are used to facilitate statistical modeling and estimation. For instance,~\cite{preda2007regression,Yuan10,cai2012minimax} use the RKHS framework in prediction problems under which the unknown slope function is assumed to reside in a reproducing kernel Hilbert space $H(K)$ with a reproducing kernel $K$, but originally the data are in $L^2$ with the usual inner product. As mentioned previously, in our case the original data are functions in a given vector-valued Reproducing Kernel Hilbert Space with the previously defined inner product, which is different from the $L^2$ inner product, and  this is an important difference between conventional functional data analysis applications and ours.

It is known that the eigen structures of integral operators in an RKHS provide bases of functions; this idea will be explored in the following section.

\subsection{Basis from the Operator integral of the Kernel}\label{sec:base-kernel}

Let $L_K\colon L^2 \longrightarrow  L^2$ be the integral operator of the kernel $K$ in the space $H_K$, defined by
\begin{equation*}
L_K f (x) :=\int_D K(x,y) (f(y)) \,dy.
\end{equation*}

Since $L_K$ is a compact, continuous, self-adjoint, positive operator, there are eigenvalues $\{\lambda_l\}_{l=1}^{\infty}$ and the corresponding eigenfunctions $\{\psi_l\}_{l=1}^{\infty}$ of $L_K$, with $\lambda_1 \geq \lambda_2 \geq \dots > 0$ and $\lim\limits_{l\rightarrow\infty} \lambda_l=0$(~\cite{Hsing15}).

Moreover,
\begin{equation*}
\langle \psi_i, \psi_j \rangle_{L^2}= \delta_{ij},\qquad \text{and}\qquad \langle\psi_i, \psi_j \rangle_{H_K}= \delta_{ij}/\lambda_i,
\end{equation*}
where $\delta_{ij}$ is the Kronecker delta.

\vspace{0.2 cm}

Therefore, if we denote $\rho_l= \sqrt{\lambda_l}\psi_l$, $\{\rho_l\}_{l=1}^{\infty}$ is an orthonormal basis for $H_K$, and the hypersurfaces $S_k$ in Eq.~\eqref{eq:funcio7}, as elements of $H_K$, can be represented as
\begin{equation}\label{eq:expresionbaseRKHS}
\overline{\varphi_k}= \sum_{l=1}^{\infty} \langle\overline{\varphi_k}, \rho_l \rangle_{H_K}\, \rho_l=\sum_{l=1}^{\infty}\mu_l^k \, \rho_l .
\end{equation}

The basis $\{\rho_l\}_{l=1}^{\infty}$ is determined by the RKHS where our functions are included and a generalization to our vector-valued case of  Theorems~4.4.7 and 4.6.8 in~\cite{Hsing15} can be proved.
Because our functional data are of the form $\overline{\varphi_k}(x)= \sum_{i=1}^N K(a_i,x)(\beta_i^k)$, these results ensure
that the truncated eigenvalue-eigen\-vector decomposition provides the best approximation to $K$ and, as a result, the truncation of Eq.~\ref{eq:expresionbaseRKHS} reduces the dimension in an optimal way.

This basis was previously used in~\cite{Barahonaetal17} in a Supervised Classification problem. As usual in practice, the coefficients are estimated using the matrix approach to the kernel function (~\cite{Barahonaetal17}).
\medskip

\subsection{Basis from the Operator integral of the Covariance function}\label{sec:base-covarianza}

Our functional data are realizations of random variables that take values in a vector-valued RKHS.
Classical $L^2$ FPCA is based on  the eigenvalue-eigenvector decomposition of the integral operator of the covariance function. We will show that this operator is also well defined in the case of random elements in vector-valued RKHS and the Karhunen-–L\'oeve Theorem is fulfilled (~\cite{Hsing15}). The eigenvectors of the decomposition of this operator are different from those of the covariance operator (see~\cite{Hsing15} page 197) and, as a result, they do not form a base of $H_K$.   We will prove that they are included in $H(K)$ and  then we will calculate the inner products in Eq.~\ref{eq:logit-final} with respect to the $H_K$ norm.

\vspace{0.2cm}

It is known that a random element of $H_K$ is a stochastic process
 (\cite{Hsing15}). Therefore, we could consider the covariance functions $\gamma_{ij}(x,y):= Cov(\Phi_i(x), \Phi_j(y))$, $\forall i,j=1, \dots, n$. Let $\gamma(x,y)$ be the $(n\times n)$-matrix whose elements are $\gamma_{ij}(x,y)$. Then  $\Gamma(x,y)(\alpha):=\gamma(x,y) \alpha$ is a symmetric and nonnegative-definite vector-valued function. We consider the integral operator $L_\Gamma$:
\begin{equation*}
L_\Gamma f (x) :=\int_D \Gamma(x,y) (f(y)) \,dy.
\end{equation*}

By again using the eigenvalue-eigenvector decomposition for a self-adjoint compact operator (~\cite{Quangetal10,Hsing15}), the eigenfunctions of the operator $L_\Gamma$, $\{v_l\}_{l=1}^{\infty}$ form an orthonormal basis for $L^2$, that is, $\langle v_i, v_j \rangle_{L^2}= \delta_{ij}$.

Because $\overline{\varphi_k} \in H_K \subset L^2$:
\begin{equation}\label{eq:base2}
\overline{\varphi_k}= \sum_{l=1}^{\infty} \langle\overline{\varphi_k}, v_l \rangle_{ L^2} v_l= \sum_{l=1}^{\infty} \varsigma^k_l v_l.
\end{equation}

As $\Gamma$ is a symmetric and non-negative-definite function, we can consider the vector-valued RKHS associated with the kernel $\Gamma$, $H_{\Gamma}$ (~\cite{Aronszajn50}).
Assuming that $\Gamma$ is continuous, it is known that $H_{\Gamma}\subset H_K$ (~\cite{Lukic01}), and then $v_j \in H_K$.

The inner products satisfy the following relationship with respect to the products in $L^2$:
\begin{equation}\label{prodvects}
\langle v_i, v_j \rangle_{H_K}= \sum_{k=1}^{\infty} \lambda_k^{-1} \langle v_i, \psi_k \rangle_{L^2} \langle v_j, \psi_k \rangle_{ L^2}.
\end{equation}

\vspace{0.2 cm}

The basis given by the integral operator defined from the covariance functions $\{v_l\}_{l=1}^{\infty}$ depends on the random sample, and the Karhunen-–L\'oeve Theorem guarantees that there are random variables $I_{\Phi}(v_l)$ with mean zero, decreasing variances and uncorrelated such that:
$$\lim_{n \rightarrow \infty} \sup_{t \in D} E[\|\Phi(t)- \sum_{l=1}^n I_{\Phi}(v_l) v_l(t)\|]=0,$$ and as a result, the truncated development of Eq.~\ref{eq:base2} is optimal in this regard.

As usual in practice, the coefficients $\varsigma^k_l$ are estimated using the matrix approach to the covariance function (~\cite{Barahonaetal17}).

This basis has some drawbacks, as mentioned in the introduction. When we use the PCA approach in logistic or multinomial functional generalized linear models, there is no guarantee that the separation between groups will be in the direction of the high-variance PCs. Moreover, following the covariance procedure, only linear relations are captured (~\cite{Gonzalez10}).

\subsection{Mixed basis}\label{sec:base-mixed}

 In the preceding subsections we have seen that we can express a function $\overline{\varphi_k}$ representing a hypersurface $S_k$ as an infinite linear combination with respect to two different bases: $\{v_l\}_{l=1}^{\infty}$, which depends on the random sample and is related to the covariance operator,  and  $\{\rho_l\}_{l=1}^{\infty}$, which is determined by the kernel that defines the RKHS. Both of them have different optimality properties.
 The aim of this section is to obtain a new basis   $\{u_l\}_{l=1}^{\infty}$, from a relationship between both operators $L_K$ and $L_\Gamma$.
 This expression will represent a compromise between the optimality given by the sample information and the one given by the RKHS in which our functions are included.

Define the linear operator $G:= L_K^{\frac12} \, \circ \, L_{\Gamma} \, \circ \, L_K^{\frac12}$ and let $\{\eta_j, w_j\}_{j=1}^{\infty}$ be the eigenvalue-eigenvector pairs of $G$.

As a consequence of Mercer's theorem, we have $\ima(L_K^{\frac12})= H_K$ (~\cite{Quangetal10}). Then, since $H_{\Gamma}\subset H_K$, we have $\ima (G)\subseteq H_K$.
\medskip

 Note that if we assume that the operators $L_K$ and $L_{\Gamma}$ are perfectly aligned, that is, they share the same ordered set of eigenfunctions $\{\psi_j\}_{j=1}^{\infty}$, then
$$G(\psi_k) = l_k \lambda_k \, \psi_k.$$

In this case, $\eta_j = l_k \lambda_k$ and the three bases  $\{\psi_j\}_{j=1}^{\infty}$, $\{v_j\}_{j=1}^{\infty}$ and  $\{w_j\}_{j=1}^{\infty}$ used in the application  coincide.

\medskip

From the basis $\{w_l\}_{l=1}^{\infty}$ we will define in Theorem~\ref{thm:simult}, a new set of functions in $H_K$, $\{u_l\}_{l=1}^{\infty}$, which generate all the functions in $H_K$, and from which we will obtain a relationship (``simultaneous diagonalization'') between the operators $L_K$ and $L_{\Gamma}$.

\begin{thm}\label{thm:simult}
Let $\{\eta_j, w_j\}_{j=1}^{\infty}$ be the eigenvalue-eigenvector pairs of $G$.\\
Define $u_j:=  L_K^{\frac12}(w_j),$  $\forall j$.\\

Then, $\forall f \in H_K $, $\ f= \displaystyle \sum_{j=1}^{\infty} \xi_j \; u_j,\ $ where $\ \xi_j = \langle f \ , \ L_K^{-1} \; u_j \rangle_{ L^2}$,
\begin{equation*}
L_{\Gamma} \; u_j= \eta_j \; L_{K}^{-1} \; u_j
\end{equation*}
and
\begin{equation*}
\langle u_i \ , \ L_{\Gamma} \; u_j \rangle_{L^2}= \eta_j\, \delta_{ij}.
\end{equation*}
\end{thm}

{\it Proof.} Details regarding operators and proof of the theorem  can be found in the appendix. $\square$
\medskip

As in the preceding subsection, the products of vectors $u_j$, which are not necessarily orthonormal with respect to the $L^2$-metric, satisfy a similar equation to Eq.~\ref{prodvects}.
\medskip

Similar bases to those used in Subsections 5.2 and 5.3 have been considered in~\cite{Yuan10} and \cite{cai2012minimax}, but in a different framework. In these papers the authors consider a prediction problem in FDA where the functional predictor is a real function defined over a domain in $\re$, and the slope function is also a real function which is assumed to reside in a real-valued RKHS. In our case, both the functional predictor and the slope are vector-valued functions in a vector-valued RKHS. Then, a metric defined in  the vector-valued RKHS is used instead of the $L^2$-metric.

\section{Application} \label{sec:our_appl}
We revisit the motivating example presented in Section~\ref{MotivatingExample} and apply the proposed modeling approach.

\begin{figure}[htbp]
\begin{center}
\includegraphics[width=6cm]{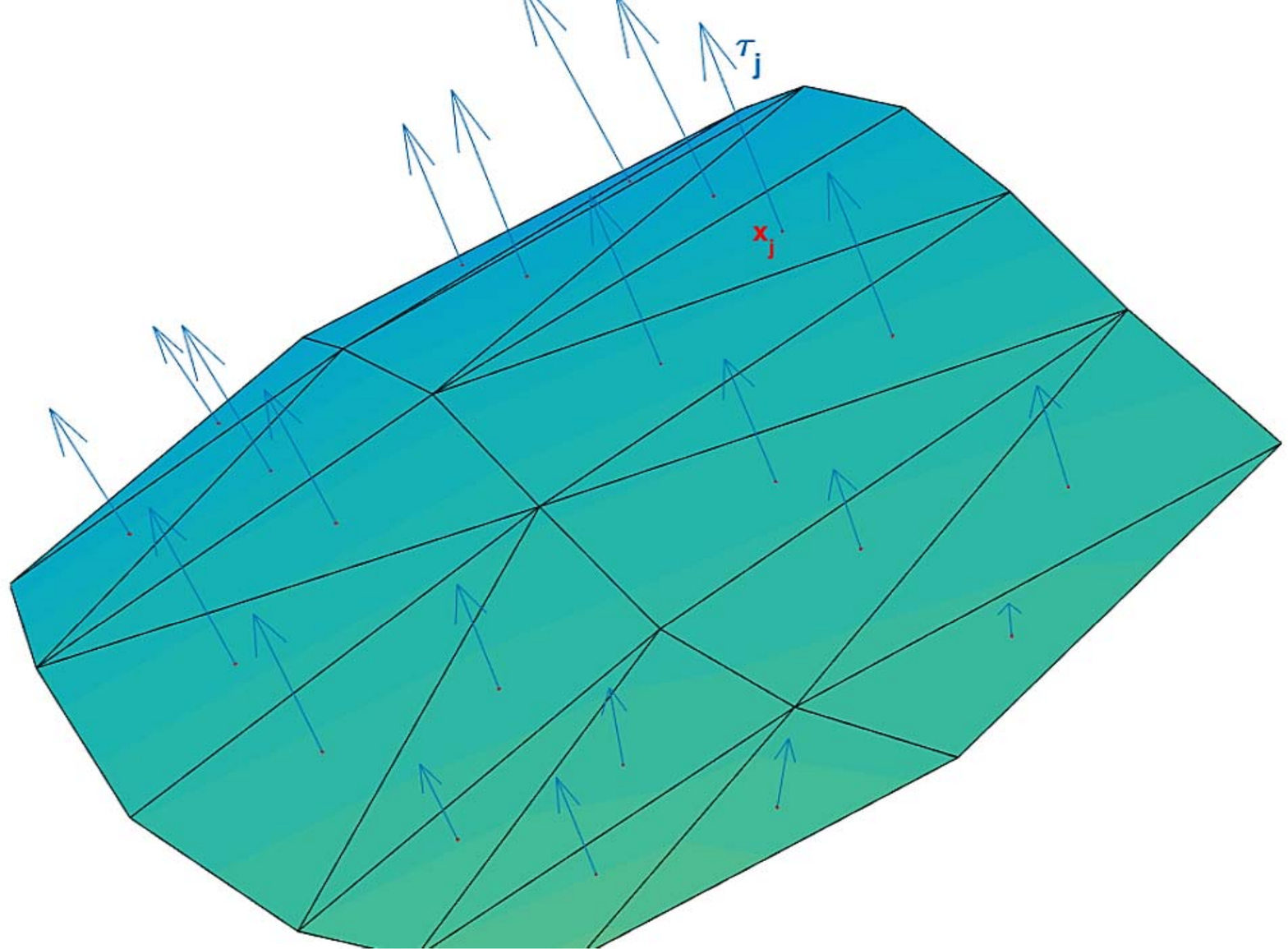}\\
\caption{ Section of the triangulated surface in $\re^3$, centers of the triangles and area vectors.
} \label{fig:triangulos}
\end{center}
\end{figure}

As stated in Section~\ref{MotivatingExample}, the points representing the surface of each child were grouped into 2766 triangles, forming a mesh.
If $a_j,b_j,c_j$ denote the vertices of the $j$-th  oriented triangle for a child $k$, the center of this triangle was defined as $x_j^k= (a_j+b_j+c_j)/3$ and its area vector (that is, its unit normal vector, scaled by its area) was $\tau_j^k= (b_j-a_j)\times(c_j-a_j)$, $\forall j=1,\dots,2766$ (see Fig.~\ref{fig:triangulos}).

\begin{figure}[htbp]
\begin{center}
 \includegraphics[width=5cm]{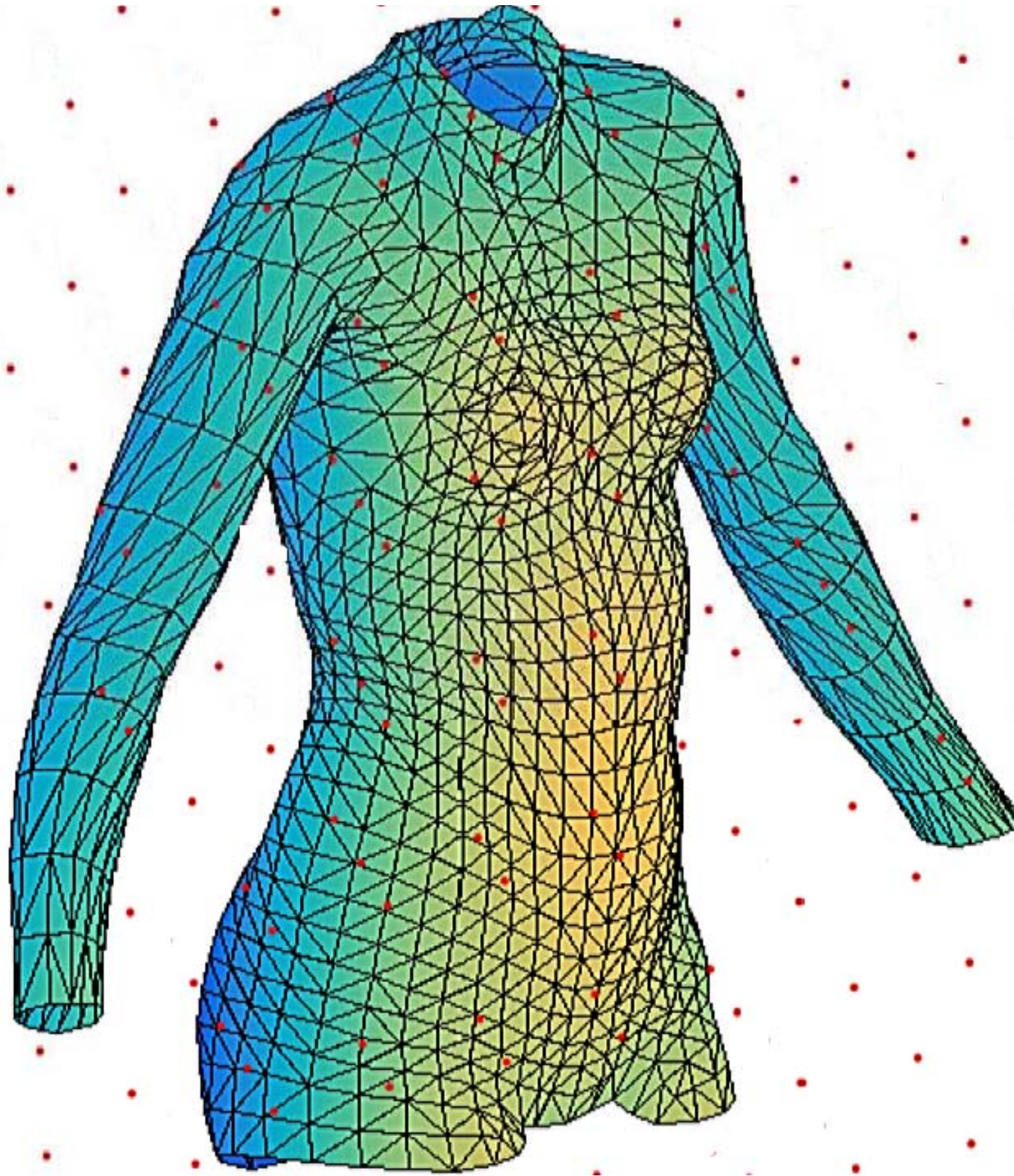} \\
\caption{Triangulated surface that represents the body contour from the upper body of a child in the sample. Red points represent the common grid $\{a_i\}_{i=1}^N$. } \label{fig:niyoenelgrid}
\end{center}
\end{figure}

Then, each child's body surface was associated with a function $\varphi_k= \sum_{j=1}^{2766} K(x_j^k, \cdot)(\tau_j^k)$ in $H_K$, where the points $x_j^k$ differ from one hypersurface to another. All these vector fields were represented in the same sample grid of points $\{a_i\}_{i=1}^N$ chosen in the compact subset $D \subset \re^3$ ,
so that each $\varphi_k$ was approximated by a smooth function $\overline{\varphi_k}=\sum_{j=1}^{N} K(a_j, \cdot)(\beta_j^k)$ evaluated on this common grid  (Sec.\ref{sec:RKHS}). In this case we considered $D =[-472.73, 487.27] \times [-824.72, 735.28] \times [-156.70, 203.30]$  and the grid was defined on this domain considering a set of points separated by a fixed gap $\Delta=200$ in the three dimensions. So a grid with $N=90$ points was obtained (see Fig.~\ref{fig:niyoenelgrid}).

A Gaussian kernel, $K(x,y):= \exp(\frac{-\norm{x-y}^2_{\re^3}}{\lambda^2})I_{3\times 3}$ was used in the definition of the operator-valued reproducing kernels $K$ (Eq.~\ref{eq:discrete}) and the value of the parameter $\lambda$ was chosen by cross-validation.

The robustness of this representation regarding the pre-processing parameters and the kernel assumption has been discussed in \citep{Barahonaetal16} and \cite{Barahonaetal17}, using a data base with much simpler geometrical objects and so with much less computational cost.

In Section~\ref{sec:base-ortonormal}, it has been seen that we can express a function $\overline{\varphi_k}$ by representing a surface $S_k$ as an infinite linear combination with respect to a basis of $H_K$ that is determined by the RKHS (Sec.~\ref{sec:base-kernel}), with respect to a  basis of $H_K$ that depends on the random sample (Sec.~\ref{sec:base-covarianza}), and with respect to a basis of $H_K$ found from both non-negative-definite operators (Sec.~\ref{sec:base-mixed}).

As our aim is to predict the goodness of fit of a given shirt size for the $k$-th child as small ($Y_k=-1$), good fit ($Y_k=0$) or large ($Y_k=1$) as a function of the garment size, the age of the child, his/her body shape and his/her sex, we will get three different models depending on the basis used to express the children's body surface in the RKHS. Following the notation used in Section~\ref{sec:FGLM}, let us denote the basis used each time with  $\{\phi_l\}_{l=1}^{\infty}$, and the corresponding coefficients for the $k$-th surface with $\{c_l^k\}_{l=1}^{\infty}$.
$\forall l=1,\dots,\infty$ we will consider
 \textit{'case 1'}, where $\phi_l= \rho_l$ and $c_l^k=\langle \overline{\varphi_k}, \rho_l \rangle_{H_K}$ (as in Eq.~\ref{eq:expresionbaseRKHS}),
 \textit{'case 2'}, where $\phi_l= v_l$ and $ c_l^k=\langle \overline{\varphi_k}, v_l \rangle_{L^2}$ (as in Eq.~\ref{eq:base2}), and
 \textit{'case 3'}, where $\phi_l= u_l$ and $ c_l^k=\langle \overline{\varphi_k}, \ L^{-1}_K \; u_l \rangle_{L^2}$ (as in Theorem~\ref{thm:simult}).
Since much of the information inherent in the original data is captured by the first few functional components and their associated coefficients, these bases are truncated with a low number $r$ of terms.
In \textit{'case 1'}, the approach $\overline{\varphi(x)}\cong \sum_{l=1}^r c_l \phi_l(x)$ is truncated considering $r = 7$ elements in the basis.  In \textit{'case 2'}, $r = 8$ elements are considered in the basis and in \textit{'case 3'} $r = 7$  elements are also considered. Then, Eq.~\eqref{eq:modelo-logit3} to include functional and non-functional predictors in the model become:
\begin{equation*}
\begin{aligned}
\logit[P(Y_k \leq j)]=& \alpha_j +\beta_1 shirt.size+\beta_2 sex_k+\beta_3 age_k+ \\
&+ \sum_{p=1}^r \sum_{l=1}^r  b_p c_l^k \langle \phi_p, \phi_l \rangle_{H_K},  \forall j\in \{-1, 0\}, \forall k
\end{aligned}
\end{equation*}
with $\{\alpha_j\}_{j\in -1, 0}$, $\beta_1$, $\beta_2$, $\beta_3$ and $\{b_p\}_{p=1}^r$, parameters to estimate.

This model assumes independent observations, but in our case several measurements are taken on each child, so this model is modified to overcome this fact into the following cumulative link mixed model~\citep{agresti10}:
\begin{equation*}
\begin{aligned}
\logit[P(Y_{k,i} \leq j)]=& \alpha_j +\beta_1 shirt.size_i+\beta_2 sex_k+\beta_3 age_k+ \\
&+ \sum_{p=1}^r \sum_{l=1}^r  b_p c_l^k \langle \phi_p, \phi_l \rangle_{H_K}+u(k),\\
&  \forall i=1,\cdots,n_k; j\in \{-1, 0\}; k=1,\cdots, 78,
\end{aligned}
\end{equation*}
where $n_k \in \{1,2,3\}$ is the number of observations taken on the $k$-th child, and the child's effects are assumed to be random, independent and identically distributed following a Gaussian distribution, i.e.
$ u(k)\sim N(0, \sigma^2_k)$.

The \textit{clmm} function of the R-package \textit{ordinal} (~\cite{ordinal}) is used to fit the model in all three cases.

Additionally, a leave-one-out cross-validation (CV) analysis is performed to check the predictive power of the model for the different bases. For each basis, the model is iteratively estimated taking into account all the data except for the observations available for each child in turn and a prediction is made for the different observations of this child (i.e. the model selection is performed within the CV and the predictions are completely out-of-sample). A percentage of agreement between predictions and real observations is computed and used to evaluate the different cases (see table~\ref{tbl:resultados}).

\begin{table}[htbp]
\begin{center}
\begin{tabular}{cc|ccc|ccc|ccc|}
\cline{3-11}
&&\multicolumn{3}{c|}{Case 1} &\multicolumn{3}{c|}{Case 2} &\multicolumn{3}{c|}{Case 3}\\
\cline{3-11}
&&\multicolumn{3}{c|}{Prediction} &\multicolumn{3}{c|}{Prediction} &\multicolumn{3}{c|}{Prediction} \\
\cline{3-11}
&& -1 & 0 & 1 & -1 & 0 & 1& -1 & 0 & 1 \\
\hline
\multicolumn{1}{|c|}{\multirowcell{3}{Expert\\decision}}
                       & -1&  47 &16   &1   &   51&12  &1  &  51 &12  &1    \\
\multicolumn{1}{|c|}{} &  0&    10&41   &10   &   11&38 &12  &  10 &39  &12    \\
\multicolumn{1}{|c|}{} &  1&    1&12   &54   &   0&13  &54  &   1&11  &55  \\
\hline
\multicolumn{1}{|c}{$\%$ of agreement} &&\multicolumn{3}{c|}{$ 73.95\%$} & \multicolumn{3}{c|}{$ 74.48\%$}&\multicolumn{3}{c|}{$ 75.52 \%$}  \\
\hline
\end{tabular}
\caption{Results of the cross-validation procedure for the mixed ordinal regression models estimated with the different bases.}
\label{tbl:resultados}
\end{center}
\end{table}

As can be seen, although in the functional data analysis literature, the basis of eigenfunctions of the covariance function is one of the most commonly used to approach functional observations, and to work with them, we have found two additional bases, and we have obtained a similar predictive power for this particular application with all of them. Although the differences are quite small, the results are slightly better with the mixed basis, as initially expected.

\subsection{Robustness against the number of points in the grid}
To evaluate the influence of the gap $\Delta$ that determines the number of points in the grid, we have repeated the previous procedure for two additional values: $\Delta=100$ and $\Delta=250$. The experimental results shown in Table~\ref{tbl:resultados2} reflect that the procedure is robust enough and the fact that better results are obtained from a balance between small and large values of $\Delta$.
\begin{table}[htbp]
\begin{center}
\begin{tabular}{|c|c|c|c|c|}
\hline
\multicolumn{1}{|c|}{\multirowcell{2}{$\Delta$}}&\multicolumn{1}{|c|}{\multirowcell{2}{Points \\in the grid }}&\multicolumn{3}{c|}{$\%$ of agreement}\\
\cline{3-5}
&& Case 1& Case 2& Case 3\\
\hline
100	&450&	71.88\%	& 69.79\% &	71.35\% \\
200	&90&	73.96\%	& 74.48\%	&75.52\% \\
250	&60&	70.31\%	&69.79\%	&70.83\% \\
\hline
\end{tabular}
\caption{Results of the cross-validation procedure for different values of $\Delta$.}
\label{tbl:resultados2}
\end{center}
\end{table}

\subsection{Comparison with other methods}

The European Standard CEN - EN 13402-3, establishes tables for body measurements and intervals to be used for compiling standard garment sizes for men, women, boys, girls and infants. Additionally, each brand usually has its own sizing chart that relates consecutive ranges of the main anthropometric
measurements with size assignations. So, several new technologies and online services have been developed in recent
years to address the selection of proper garment sizes or models based on the user's anthropometric measurements (see, for instance, \url{www.fits.me}).
In the case of T-shirts, shirts and/or blouses, the main anthropometric dimensions established by the European standard  are height, chest and neck for boys, and height and bust for girls.

In~\cite{pierolaetal16}, the authors used ordered logistic regression and random forest methodologies to predict a garment's goodness of  fit from the differences between the measurements of the reference mannequin for the evaluated size and the child's anthropometric measurements.

Following this line, we could also have  used different children's anthropometric measurements to fit a mixed proportional odds model, as in~\cite{Mccullagh80}. So if $Y_{k,i}$ denotes the response as (small (-1)/good (0)/large (1)) fit of a shirt size $i$ for the $k-th$ child, and $X_k$ denotes a vector of explicative variables formed by the sex of the $k-th$ child, his/her age and the values of the 27 anthropometric measurements considered by~\cite{pierolaetal16}, we could have fitted:

\begin{eqnarray}\label{mixedorderedlogistico}
&& \logit[P(Y_{k,i} \leq j)]=\alpha_j +\beta_1 shirt.size_i+\beta X_k+u(k), \\
&& \forall i=1,\cdots,n_k; j\in \{-1, 0\}; k=1,\cdots, 78, \nonumber
\end{eqnarray}
where once again $n_k \in \{1,2,3\}$ is the number of observations taken on the $k$-th child, and the child's effects are assumed to be random, independent and identically distributed following a Gaussian distribution, i.e. $ u(k)\sim N(0, \sigma^2_k)$.

Performing a leave-one-out cross-validation study, choosing the
model on each step  by a forward stepwise model selection based on likelihood ratio tests (~\cite{ordinal}), we obtain worse results than those obtained with our methodology. The percentage of correct classifications is now $68.27\%$ (see table \ref{tbl:resultados2}).

As in our case, many problems in medical imaging analysis and computer vision involve the classification of bodies (geometrical objects with bounded
boundaries), based  on their size and shape. Several mathematical frameworks have been proposed in the literature to deal with such
objects, three of these being the most widely used. Firstly, functions can be used to represent closed
contours of the objects (curves in 2D and surfaces in 3D). Secondly, geometrical objects can be treated as compact subsets of $\re^m$ and, finally, these geometrical objects can be characterized by sequences of points with certain geometrical or anatomical properties (landmarks).

In our application, we are working with rigidly aligned 3D homologous avatars with anatomical one-to-one vertex correspondence among them (~\cite{ballesteretal14}), so we can consider these 1423 points as landmarks, and define $X_{k} \in M_{1423\times 3}$ as the configuration matrix of the $k$-th child. As the shape of an object is all the geometric information that remains invariant with translations, rotations and changes of scale, the shape space and size-and-shape space are not flat Euclidean spaces, so  classical statistical methods cannot be directly applied to the manifold-valued data (~\cite{Pennec06}). However, if the sample has little variability, the problem can be transferred to a tangent space (at the Procrustes mean of these shapes or size-and-shapes, for example) and then standard multivariate procedures can be performed in this space (~\cite{dryden2016statistical}), such as Principal Component Analysis (PCA), where the first $p$ PC scores, which summarize most of the variability in the tangent plane data, are usually chosen in order to reduce the dimensionality of the data set.

The tangent space is defined from a point called pole, so  the distance from the shape to the pole is preserved. As one moves away from the pole, the Euclidean distances between some pairs of points in the tangent space are smaller than their corresponding shape distances. This distortion becomes larger as one considers points further from it. For this reason, the pole should be taken close to all of the points and the mean of the observed shapes is the best choice (~\cite{dryden2016statistical}).

So, given the configuration matrices $X_{k} \in M_{1423\times 3}$, the size $s_k$ of each child is obtained and the full Procrustes mean shape is computed. Then, the coordinates of the projection of $X_{k} \in M_{1423\times 3}$ onto the tangent plane defined at its corresponding mean shape is obtained. The first PC scores of these coordinates are calculated and they will be used as covariates in our predictive model. The first PC components that explain $98 \%$ of the variability are considered.

So, given the response variable $Y$ with $3$ ordered categories and given the garment size to evaluate and a vector $X$ with the child's size, his/her sex, his/her age and the first PC scores of his/her coordinates in the tangent space, we can fit the model given by Eq.~\ref{mixedorderedlogistico}.

Once again, performing a leave-one-out cross-validation study using this model, we obtain worse results than those obtained with our methodology. The percentage of correct classification is now  $66.67$ (see table \ref{tbl:resultados2}).

\begin{table}[htbp]
\begin{center}
\begin{tabular}{cc|ccc|ccc|}
\cline{3-8}
&&\multicolumn{3}{c|}{Multivariate} &\multicolumn{3}{c|}{Landmarks} \\
\cline{3-8}
&&\multicolumn{3}{c|}{Prediction} &\multicolumn{3}{c|}{Prediction} \\
\cline{3-8}
&& -1 & 0 & 1 & -1 & 0 & 1 \\
\hline
\multicolumn{1}{|c|}{\multirowcell{3}{Expert\\decision}}
                       & -1&  47 &12   &3   &   48&14  &2    \\
\multicolumn{1}{|c|}{} &  0&   13&31   &15   &   14&34 &13   \\
\multicolumn{1}{|c|}{} &  1&    1&15   &49   &   0&21  &46   \\
\hline
\multicolumn{1}{|c}{$\%$ of agreement} &&\multicolumn{3}{c|}{$ 68.27\%$} & \multicolumn{3}{c|}{$ 66.67\%$} \\
\hline
\end{tabular}
\caption{Results of the cross validation procedure for the two alternative methods tested}
\label{tbl:resultados2}
\end{center}
\end{table}
\section{Discussion} \label{sec:conclusions}

In this paper we have proposed a new methodology for modeling an ordinal response variable in terms of 3D geometrical objects. It is based on their characterization by means of currents and the expression of each geometrical object in terms of three different bases of functions that generate the corresponding vector-valued RKHS.

Firstly, the predictors were expressed in the orthonormal basis given by the kernel of the RKHS. Secondly, we used a basis obtained in a similar way as the usual basis given by the covariance kernel in the scalar setting. Thirdly, a basis of functions that connect the benefits of the two previous expressions was sought by ``simultaneous diagonalization'' of operators. The coefficients of each geometrical predictor in relation to the three bases of functions were estimated. The goodness of the method was checked by leave-one-out cross-validation.

Then, it was applied to predict whether the size fits a customer or is too large or small for him/her, which is useful for an application to online clothing sales. This was done using a 3D training database obtained from an anthropometric survey of the Spanish child population. The results were quite promising, taking into account the difficulty of the application.  Although the results obtained are quite similar with the three bases tested, as expected they are slightly better with the mixed basis. In near future this methodology could be incorporated into the mobile application (kidsize) recently developed by  the Biomechanical  Institute of Valencia.

We compare our methodology with another two methods traditionally used in  biometric size determination. The first is based on considering children's anthropometric  measurements and classical ordinal regression. The second is based on landmark configuration and its projection in the tangent space where classical multivariate statistical methods can be applied. In both cases, the classification results obtained are slightly worse than those obtained with the methodology developed in this paper.

It is important to note that the success of the three bases proposed in our methodology depends on the data and the application, so our suggestion to practitioners would be to check the performance of all of them for each particular problem.

\appendix
\section{Theorem 5.1} \label{app}
Before starting with the proof of the Theorem~\ref{thm:simult}, let us recall some definitions regarding operators. If an operator ${\cal F}$ is nonnegative, there is a unique nonnegative operator ${\cal F}^{\frac12}$, called the square-root of the operator, such that $({\cal F}^{\frac12})^2= {\cal F}^{\frac12}\circ {\cal F}^{\frac12}= {\cal F}$, and it commutes with any operator that commutes with ${\cal F}$. When an operator ${\cal F}$ is bijective, there is an operator ${\cal F}^{-1}$, called the inverse operator, such that ${\cal F}^{-1}\circ {\cal F}$ and ${\cal F}\circ {\cal F}^{-1}$ are the identity operator.\\

Due the fact that $L_K$ is compact, positive-definite and self-adjoint, the operator $L_S^{\frac12}\colon L^2 \longrightarrow L^2$ can be defined, and it works as follows:
\begin{equation*}
L_K^{\frac12} (f) :=\sum_{j=1}^{\infty} \sqrt{\lambda_{j}} \langle f, \psi_j \rangle_{H_K} \psi_j.
\end{equation*}

$L_K^{\frac12}$ is also compact, positive-definite and self-adjoint.

Note that $L_K$ and $L_K^{\frac12}$ are injective because they are positive-definite (their eigenvalues are strictly positive). Hence, restricting their arrival spaces to their respective images converts them into bijective operators on pre-Hilbert spaces. These restricted operators will also be denoted by  $L_K$ and $L_K^{\frac12}$; then, there are the inverse operators o $L_{K}^{-1}$, $L_{K}^{-\frac12}$, and we have  $L_{K}^{-\frac12} \circ L_{K}^{-\frac12} \; f = L_{K}^{-1} \; f, \ \forall f \in  H_K.$
\medskip

{\it Proof of Theorem~\ref{thm:simult}.}
\vspace{0.2 cm}

Since $u_j:=  L_K^{\frac12}(w_j)$  and $G \, w_j= \eta_j \, w_j$, we prove that, for all $j$.
\begin{equation*}
\begin{aligned}
L_{\Gamma} \; u_j & =\ L_{\Gamma} \; L_K^{\frac12} \; w_j = L_{K}^{-\frac12} \; L_K^{\frac12} L_{\Gamma} \; L_K^{\frac12} \; w_j =\ L_{K}^{-\frac12} \;  G \; w_j \\
&=  L_{K}^{-\frac12} \; \eta_j \; w_j =\eta_j \; L_{K}^{-\frac12} \; L_{K}^{-\frac12} \; u_j =\eta_j \; L_{K}^{-1} \; u_j.
\end{aligned}
\end{equation*}

\vspace{0.2 cm}

Moreover, $L_K^{\frac12}$ is self-adjoint and $\{w_j\}_{j=1}^{\infty}$ is an orthogonal system. Thus,
\begin{equation*}
\begin{aligned}
\langle u_i \ , \ L_{\Gamma} \; u_j \rangle_{L^2} &= \ \langle L_K^{\frac12}\; w_i \ , \ L_{\Gamma} \;L_K^{\frac12}\; w_j \rangle_{L^2}
=  \langle w_i \ , \ L_K^{\frac12}\; L_{\Gamma} \;L_K^{\frac12}\; w_j \rangle_{L^2} \\
&=  \langle w_i \ , \ \eta_j \; w_j \rangle_{L^2} = \eta_i \delta_{ij}.
\end{aligned}
\end{equation*}

\vspace{0.2 cm}
Finally, if $f \in H_K$,
\begin{equation*}
\begin{aligned}
f & =L_K^{\frac12} \; \left( L_{K}^{-\frac12} \; f \right) = L_K^{\frac12} \left( \sum_{j=1}^{\infty} \langle  L_{K}^{-\frac12}\; f \ , \ w_j\rangle_{L^2} \; w_j  \right) \\
&= \sum_{j=1}^{\infty} \langle  L_{K}^{-\frac12}\; f \ , \ w_j\rangle_{L^2} \; L_K^{\frac12} \; \left( w_j \right) \\
&= \sum_{j=1}^{\infty} \langle  L_{K}^{-\frac12}\; f \ ,  \; L_{K}^{-\frac12} \; u_j \rangle_{L^2} \; L_K^{\frac12} \; \left(  L_{K}^{-\frac12} \; u_j \right) \\
&= \sum_{j=1}^{\infty} \langle  L_{K}^{-\frac12}\; f \ , \ L_{K}^{-\frac12} \; u_j \rangle_{L^2} \; u_j  = \sum_{j=1}^{\infty} \langle  f \ , \ L_{K}^{-\frac12}\; L_{K}^{-\frac12} \; u_j \rangle_{L^2} \; u_j \\
&= \sum_{j=1}^{\infty} \; \langle  f \ , \; L_{K}^{-1}\; u_j \rangle_{L^2} \; u_j = \sum_{j=1}^{\infty} \xi_j \; u_j ,
\end{aligned}
\end{equation*}
  $\{w_j\}_{j=1}^{\infty}$ being an orthogonal basis of $L^2$ with respect to the $L^2$-metric and using the definition of  $\{u_j\}_{j=1}^{\infty}$. $\square$

\end{document}